\documentclass[twocolumn,showpacs,preprintnumbers,prd,superscriptaddress,nofootinbib]{revtex4-1}
\usepackage{graphicx}
\usepackage{epsf}
\usepackage{bm}
\usepackage{amsmath}
\usepackage{amsfonts}
\usepackage{amssymb}
\usepackage{epstopdf}
\usepackage{natbib}
\usepackage{hyperref}
\usepackage{color}
\usepackage{verbatim}
\usepackage{multirow}
\usepackage{bm}
\usepackage{hyperref}
\usepackage{float}
\usepackage{xcolor}

\definecolor{darkblue}{rgb}{0.0, 0.0, 0.55}
\definecolor{darkred}{rgb}{0.55, 0.0, 0.0}

\usepackage{hyperref}
\hypersetup{
    colorlinks=true, 
    linkcolor=darkblue,
    citecolor=darkblue,
    urlcolor=darkblue}
    
\makeatletter\let\expandableinput\@@input\makeatother

\begin{document}

\title{Nonstandard Dark Matter at Nonlinear Scales}

\author{Rafael C. Nunes}
\email{rafadcnunes@gmail.com; costa.nunes@ufrgs.br}
\affiliation{Instituto de F\'{i}sica, Universidade Federal do Rio Grande do Sul, 91501-970 Porto Alegre RS, Brazil}
\affiliation{Divisão de Astrofísica, Instituto Nacional de Pesquisas Espaciais, Avenida dos Astronautas 1758, São José dos Campos, 12227-010, São Paulo, Brazil}

\begin{abstract}
We investigate how two extensions of the standard $\Lambda$CDM 
scenario affect the evolution of the matter density and velocity 
divergence fields, from the linear regime to deeply nonlinear 
scales, using $N$-body simulations. We consider two 
well-motivated scenarios in which the canonical properties of 
dark matter (DM) are modified. In the first scenario, DM 
interacts with dark energy (DM--DE interaction). In the second 
scenario, we allow the DM equation-of-state parameter to deviate 
from the canonical pressureless limit, $w_{\rm DM}=0$. For the 
coupled DM--DE scenario, we find significant and scale-dependent 
deviations from $\Lambda$CDM of order $50$--$100\%$ in the 
matter power spectrum, persisting throughout the nonlinear 
regime, with the velocity divergence spectra providing 
complementary and partially independent sensitivity. For the 
$w_{\rm DM} \neq 0$ scenario, we find systematic but 
remarkably small deviations confined within $5\%$ across 
all scales probed. Our results identify which scales carry the 
greatest discriminatory power between these nonstandard 
scenarios and $\Lambda$CDM. We show that testing these models 
with current and forthcoming LSS surveys, 
through the analysis of density and/or velocity fields, 
will require a deeper theoretical understanding and accurate 
nonlinear modelling before robust constraints can be derived 
from observational data.
\end{abstract}

\maketitle

\section{Introduction}

Providing a physical explanation for the origin and nature of the dark sector of the Universe, comprising dark matter (DM) and dark energy (DE), remains one of the major challenges in contemporary cosmology. The $\Lambda$CDM model, which provides the standard theoretical framework for modern cosmology, has successfully accounted for a broad range of cosmological observations over the past two decades. Nevertheless, the increasing precision of cosmological measurements has revealed several observational discrepancies that have motivated renewed scrutiny of the standard cosmological paradigm. Among these, the tensions in the inferred values of the Hubble constant, $H_0$, and the amplitude of matter clustering, commonly quantified by $S_8$, have attracted particular attention and remain at the forefront of current cosmological research (see, e.g., Ref.~\cite{CosmoVerseNetwork:2025alb,Pantos:2026koc,Hu:2023jqc,Cai:2026swf} for a review).

Recent results from the Dark Energy Spectroscopic Instrument (DESI) have further intensified this discussion by providing new constraints on the expansion history of the Universe through measurements of baryon acoustic oscillations (BAO)~\cite{DESI:2025zgx, DESI:2025fii}. In particular, analyses combining the DESI BAO measurements with different cosmic microwave background (CMB) data sets and Type Ia supernova (SNe Ia) samples have yielded preferences for evolving dark energy over a cosmological constant, with the statistical significance depending on the combination of data sets and the adopted analysis framework. Although these results do not yet provide conclusive evidence for a departure from $\Lambda$CDM, they motivate further investigation of extensions to the standard cosmological model. In particular, they raise the possibility that the observed cosmic acceleration may be driven by a dynamical DE component and/or that the DM sector may exhibit nonstandard properties~\cite{Li:2026asg,Silva:2025twg,Ladeira:2026jne,Montefalcone:2026iga,Xu:2026sbw,Li:2026xaz,deCruzPerez:2025dni,Chudaykin:2025lww,Efstratiou:2025iqi,Smith:2025icl, Li:2025muv,Liu:2025bss,RoyChoudhury:2025iis,Fazzari:2025lzd,Chaudhary:2025vzy,Gomez-Valent:2025mfl,Guedezounme:2025wav,Sabogal:2025jbo,Giare:2026gdn,Li:2026jgg,Kessler:2026dbi,Gonzalez-Fuentes:2026rgu,Sahlu:2026bsa,Akarsu:2026anp,Capozziello:2025qmh,Wolf:2025acj,Lee:2025pzo,Mishra:2025goj,Akarsu:2024qsi,Garcia-Garcia:2026nzy,Reboucas:2026ljj,Giare:2026tyk,Qiang:2025cxp,Montefalcone:2026iga,Lu:2025sjg}.

In this context, large-scale structure (LSS) surveys play a crucial role in testing phenomenological extensions of the $\Lambda$CDM model, providing complementary and independent information on the expansion history of the Universe and the growth of cosmic structures. These observations offer powerful avenues for testing departures from the standard cosmological paradigm and constraining the physical properties of the dark sector. However, interpreting LSS observations beyond the linear regime remains a major theoretical challenge, as nonlinear gravitational evolution introduces complex mode coupling and scale-dependent effects that cannot be adequately described by linear perturbation theory alone. Several theoretical approaches have been developed to address this problem, including the effective field theory of large-scale structure (EFTofLSS)~\cite{Baumann:2010tm,Carrasco:2013mua,Carrasco:2012cv,Ivanov:2026nlf}, halo models~\cite{Asgari:2023mej,Mead:2020vgs}, and Lagrangian perturbation theory and related Lagrangian approaches~\cite{Vlah:2014nta, Matsubara:2015ipa, Matsubara:2007wj,Chen:2020fxs,Vlah:2015sea}.

The matter power spectrum serves as a fundamental statistical probe connecting theoretical models to observations of large-scale structure (LSS), enabling the characterization of the cosmic matter distribution over a broad range of scales. It carries valuable information about the initial density perturbations, their subsequent gravitational evolution, and the impact of the underlying cosmological framework on the formation of structures. In particular, the nonlinear matter power spectrum plays a key role in the analysis of ongoing and future LSS surveys, describing the complex evolution of matter clustering on scales where linear perturbation theory is no longer applicable. Consequently, reliable theoretical modeling of the nonlinear power spectrum is essential for obtaining robust constraints on cosmological parameters from observational data and identifying potential signatures of physics beyond the standard $\Lambda$CDM model.

In this work, we perform $N$-body simulations to investigate the evolution of the density contrast, $\delta$, and the velocity divergence, $\theta$, in two classes of nonstandard dark matter scenarios. First, we consider a well-motivated class of interacting dark sector models in which energy-momentum is exchanged between dark matter (DM) and dark energy (DE) through a non-gravitational coupling. This extension of the standard $\Lambda$CDM cosmology has received considerable attention in the literature and has been extensively investigated using a variety of cosmological observations (e.g., Refs.~\cite{Silva:2025hxw,Wang:2026crv,Giare:2026gdn,Yang:2026qon,Li:2026jgg,Bagherian:2026jxz,Sabogal:2026bsw,Zhang:2026dtd,deSouza:2026qyf,Khoury:2026svx,Wang:2026gvg,Chanda:2026dmx,Meng:2026lzz,KumarSharma:2026ppi,Ambelu:2026mkd,Paliathanasis:2026acn,Paliathanasis:2026ymi,Lyu:2025nsd,vanderWesthuizen:2025mnw,vanderWesthuizen:2025vcb,Yang:2025uyv,Wang:2025znm,Yashiki:2025loj,Duchaniya:2025kok,Pan:2025qwy,SanchezLopez:2025uzw,Biviano:2026wur,KumarSharma:2026ppi,Figueruelo:2026eis,Sahoo:2025cvz,You:2025uon,Zhai:2025hfi,Chakraborty:2025syu,Saridakis:2026ybq}). Previous studies have also explored the matter power spectrum, $P(k)$, in interacting dark sector scenarios, including its behavior in the nonlinear regime (see, e.g., Refs.~\cite{Silva:2025bnn,Zhai:2026uwr,Zhang:2018mlj,Zhao:2022ycr,Baldi:2011qi,Silva:2024ift,Wolney:2026zbc}). Here, we focus on a specific subclass of these models that has attracted particular interest in recent literature and has been extensively tested using geometric cosmological probes, while its perturbative evolution has also been investigated in the linear regime (see, e.g., Ref.~\cite{Silva:2025hxw}). Our goal is to extend this investigation into the nonlinear regime and assess the impact of the dark sector coupling on the evolution of the density and velocity fields.

In the second part of our analysis, we investigate a complementary scenario in which the DM equation-of-state parameter deviates from the standard pressureless limit, $w_{\rm DM}=0$. Such departures from the conventional cold dark matter description have also been widely discussed and constrained using different cosmological observations (see, e.g., Refs.~\cite{Kumar:2019gfl, Yang:2025ume,Yao:2025twv, Li:2025dwz, Yao:2025kuz,Wang:2025zri,Yao:2023ybs,Ilic:2020onu,Sakr:2026jux,Abedin:2025dis}). Here, we focus on their implications for nonlinear structure formation, examining how a nonzero DM equation-of-state parameter affects the matter power spectra associated with the density and velocity fields.

Throughout this work, we retain the standard notation adopted in the literature, where the subscript $c$ denotes cold dark matter in coupled DM–DE models, while the term DM is used more generally to refer to dark matter.

The remainder of this paper is organized as follows. In Sec.~\ref{model}, we introduce the theoretical framework and describe the cosmological scenarios considered in this work. In Sec.~\ref{methodology}, we present the numerical methods and $N$-body simulation setup. In Sec.~\ref{results}, we discuss our results for the density and velocity power spectra in the nonlinear regime. Finally, in Sec.~\ref{final}, we summarize our main findings and present our conclusions.

\section{Theoretical Framework}
\label{model}

In this section, we briefly review the theoretical framework and the relevant cosmological quantities required to implement the models considered in this work in $N$-body simulations. We summarize the key equations governing the background expansion and the evolution of density and velocity perturbations, highlighting their dependence on the underlying cosmological scenario. These quantities provide the necessary theoretical input for consistently incorporating departures from the standard $\Lambda$CDM framework into our numerical simulations.

\subsection{Coupled DM--DE}

The interaction between DM and DE is phenomenologically 
described by a coupling between their stress-energy tensors, 
whose covariant divergences read
\begin{align}
\nabla_\nu T_c^{\mu\nu} &= Q\, u_c^\mu,
\label{eq:DM_conservation}
\\
\nabla_\nu T_x^{\mu\nu} &= -Q\, u_c^\mu,
\label{eq:DE_conservation}
\end{align}
where $u_c^\mu$ is the DM four-velocity and $Q$ denotes
the interaction rate. The total stress-energy is conserved
by construction. We adopt the widely used prescription
$Q = \xi\,\mathcal{H}\,\rho_x$, where $\mathcal{H} = aH$ is the conformal Hubble rate,
$\rho_x$ is the DE energy density, and $\xi$ is a
dimensionless coupling parameter. Positive (negative) $\xi$
describes energy transfer from DE to DM (DM to DE).

In conformal time $\tau$, the continuity equations for the
two dark components are
\begin{align}
\dot\rho_c + 3\mathcal{H}\rho_c
    &= \xi\mathcal{H}\rho_x,
\label{eq:DM_continuity}
\\
\dot\rho_x + 3\mathcal{H}(1+w_x)\rho_x
    &= -\xi\mathcal{H}\rho_x,
\label{eq:DE_continuity}
\end{align}
where a dot denotes $d/d\tau$ and $w_x \equiv p_x/\rho_x$
is the DE equation-of-state parameter, assumed constant.

The coupling induces an effective equation of state for
pressureless DM,
\begin{equation}
w_{\rm DM,eff}(a) = -\frac{\xi}{3}\,\frac{\rho_x(a)}{\rho_c(a)},
\label{eq:weff}
\end{equation}
reflecting the energy exchange rather than any intrinsic DM
pressure.

Working in the Newtonian gauge, the coupled
Einstein--Boltzmann equations governing the DM density
contrast $\delta_c$ and velocity divergence $\theta_c$
are~\cite{Valiviita:2008iv,Gavela:2010tm}
\begin{align}
\dot\delta_c &= -\left(\theta_c - 3\dot\Phi\right)
    + \frac{Q}{\rho_c}
      \left(\frac{\delta Q}{Q} - \delta_c + \Psi\right),
\label{eq:delta_c}
\\
\dot\theta_c &= -\mathcal{H}\theta_c + k^2\Psi,
\label{eq:theta_c}
\end{align}
where dots denote $d/d\tau$. In the Newtonian limit on
sub-horizon scales, DE perturbations are negligible, so that $\delta Q/Q \approx \delta_x \approx 0$.
%
%
The Euler equation~(\ref{eq:theta_c}) is unaltered with
respect to $\Lambda$CDM, confirming the absence of a fifth
force in this model. The Poisson equation retains its
standard form,
\begin{equation}
-k^2\Psi = 4\pi G a^2\rho_c\,\delta_c,
\label{eq:Poisson}
\end{equation}
The linear growth factor $D(a) \equiv \delta_c(a)/\delta_c(1)$
is obtained by differentiating equation~(\ref{eq:delta_c})
with respect to $\tau$, substituting $\dot\theta_c$ from
equation~(\ref{eq:theta_c}) and $k^2\Psi =
-(3/2)\mathcal{H}^2\Omega_c\delta_c$ from
equation~(\ref{eq:Poisson}), which gives
\begin{equation}
\begin{split}
\ddot\delta_c
    &+ \mathcal{H}\left(1 + \xi r_{xc}\right)\dot\delta_c \\
    &- \left[\frac{3}{2}\mathcal{H}^2\Omega_c
      \left(1 + \xi r_{xc}\right)
      - \xi\dot{\mathcal{H}}r_{xc}
    \right]\delta_c = 0.
\end{split}
\label{eq:growth_conformal_IDE}
\end{equation}
Converting to derivatives with respect to $\ln a$
via $\dot f = \mathcal{H}f'$,
$\ddot f = \mathcal{H}^2(f'' + \mathcal{H}'/\mathcal{H}\cdot f')$
and $\dot{\mathcal{H}} = \mathcal{H}'\mathcal{H}$,
and dividing through by $\mathcal{H}^2$, yields
\begin{equation}
\begin{split}
D'' &+ D'\!\left[
    \frac{\mathcal{H}'}{\mathcal{H}} + 1
    + \xi\,r_{xc}\right] \\
    &- D\!\left[
    \frac{3}{2}\Omega_c\!\left(1 + \xi\,r_{xc}\right)
    - \frac{\mathcal{H}'}{\mathcal{H}}\,\xi\,r_{xc}
    \right] = 0,
\end{split}
\label{eq:growth}
\end{equation}
where primes denote $d/d\ln a$ and
$r_{xc} \equiv \rho_x/\rho_c$. For $\xi = 0$,
the coefficient of $D'$ reduces to
$\mathcal{H}'/\mathcal{H} + 1$ and the source term
reduces to $(3/2)\Omega_c = (3/2)\Omega_m$ ((neglecting the
small baryon contribution to the growth equation),
recovering the standard $\Lambda$CDM growth equation.
The growth factor is normalised to unity today,
$D(1)=1$, and enters the initial conditions of the
$N$-body simulations via the Zel'dovich approximation.

The velocity divergence field
$\theta_c \equiv \nabla\cdot\mathbf{v}_c$ provides
complementary information to the density field. In linear
theory, it is related to the growth factor through the
continuity equation~\cite{Bernardeau:2001qr},
\begin{equation}
\theta_c = -\mathcal{H}\,f\,\delta_c,
\label{eq:theta_linear}
\end{equation}
where $f \equiv d\ln D/d\ln a$ is the linear growth rate,
obtained directly from equation~(\ref{eq:growth}). In the
IDE framework, $f$ is sensitive to the coupling $\xi$
both through the modified background $H(a)$ and through
the additional terms in the growth equation. In the limit
$\xi = 0$, the standard approximation is recovered.

The associated linear power spectra are
\begin{align}
P_{\theta\theta}(k,a) &=
    \left[f(a)H(a)\right]^2 P_{\delta\delta}(k,a),
\label{eq:Ptt}
\\
P_{\delta\theta}(k,a) &=
    f(a)H(a)\,P_{\delta\delta}(k,a),
\label{eq:Pdt}
\end{align}
where $H = \mathcal{H}/a$ is the cosmic-time Hubble rate.
These relations break down at non-linear scales, where
shell-crossing and vorticity generation decouple
$\theta$ from $\delta$~\cite{Pueblas:2008uv}.

\subsection{Dark Matter with Non-Zero Equation of State}
\label{wDM}

In the second scenario we consider, we relax the
canonical assumption of pressureless dark matter and
allow the DM equation-of-state parameter $w_{\rm DM}$
to deviate from zero. This represents an intrinsic
modification of the DM fluid, distinct from the
effective pressure induced by the IDE coupling. A non-zero $w_{\rm DM}$ can arise in several well-motivated frameworks (e.g.,~\cite{Kumar:2019gfl,Blanchet:2024mvy,Skordis:2020eui,Kopp:2018zxp,Saridakis:2026lzb}).

For a DM fluid with constant equation of state
$w_{\rm DM}$, the continuity equation reads
\begin{equation}
\dot\rho_{\rm DM} + 3\mathcal{H}(1 + w_{\rm DM})
    \rho_{\rm DM} = 0,
\label{eq:wDM_continuity}
\end{equation}
where a dot denotes $d/d\tau$, which integrates to
\begin{equation}
\rho_{\rm DM}(a) = \rho_{\rm DM,0}\,
    a^{-3(1+w_{\rm DM})},
\label{eq:rho_wDM}
\end{equation}
where for $w_{\rm DM} = 0$ the standard $\Lambda$CDM
expression is recovered.

At the perturbation level, a non-zero $w_{\rm DM}$
modifies both the continuity and Euler equations.
Working in the Newtonian gauge, where a dot denotes
$d/d\tau$~\cite{Kumar:2019gfl}:
\begin{align}
\dot\delta_{\rm DM} &= -(1 + w_{\rm DM})
    \left(\theta_{\rm DM} - 3\dot\Phi\right)
    - 3\mathcal{H}\left(c_s^2 - w_{\rm DM}\right)
    \delta_{\rm DM},
\label{eq:delta_wDM}
\\
\dot\theta_{\rm DM} &= -\mathcal{H}
    (1 - 3c_s^2)\theta_{\rm DM}
    + \frac{c_s^2 k^2}{1+w_{\rm DM}}\delta_{\rm DM}
    + k^2\Psi,
\label{eq:theta_wDM}
\end{align}
where $c_s^2$ is the rest-frame sound speed of the
DM fluid. The Poisson equation retains its standard
form, equation~(\ref{eq:Poisson}).

To obtain the linear growth equation for
$D(a) \equiv \delta_{\rm DM}(a)/\delta_{\rm DM}(1)$,
we differentiate equation~(\ref{eq:delta_wDM}) with
respect to $\tau$, neglecting $\dot\Phi$
on sub-horizon scales:
\begin{equation}
\begin{split}
\ddot\delta_{\rm DM} &=
    -(1+w_{\rm DM})\dot\theta_{\rm DM} \\
    &\quad - 3\dot{\mathcal{H}}(c_s^2-w_{\rm DM})
      \delta_{\rm DM} \\
    &\quad - 3\mathcal{H}(c_s^2-w_{\rm DM})
      \dot\delta_{\rm DM}.
\end{split}
\label{eq:ddot_delta}
\end{equation}
Substituting $\dot\theta_{\rm DM}$ from
equation~(\ref{eq:theta_wDM}) and the Poisson equation
$k^2\Psi = -4\pi G a^2\rho_{\rm DM}\delta_{\rm DM}
= -(3/2)\mathcal{H}^2\Omega_{\rm DM}\delta_{\rm DM}$,
equation~(\ref{eq:ddot_delta}) becomes
\begin{equation}
\begin{split}
\ddot\delta_{\rm DM}
    &+ \mathcal{H}(1 - 3w_{\rm DM})
      \dot\delta_{\rm DM} \\
    &- \left[\frac{3}{2}(1+w_{\rm DM})(1+3c_s^2)
      \mathcal{H}^2\Omega_{\rm DM}(a)
      - c_s^2 k^2\right]\delta_{\rm DM} \\
    &- 3\dot{\mathcal{H}}(c_s^2-w_{\rm DM})
      \delta_{\rm DM} = 0,
\end{split}
\label{eq:growth_conformal}
\end{equation}
where the coefficient of $\dot\delta_{\rm DM}$
follows from $(1-3c_s^2) + 3(c_s^2-w_{\rm DM})
= 1 - 3w_{\rm DM}$.

We now convert to derivatives with respect to
$\ln a$, using $\dot{f} = \mathcal{H}f'$ and
$\ddot{f} = \mathcal{H}^2(f'' +
\mathcal{H}'/\mathcal{H}\cdot f')$,
together with $\dot{\mathcal{H}} =
\mathcal{H}'\mathcal{H}$, so that
$3\dot{\mathcal{H}}(c_s^2-w_{\rm DM})/\mathcal{H}^2
= 3(\mathcal{H}'/\mathcal{H})(c_s^2-w_{\rm DM})$.
Dividing through by $\mathcal{H}^2$ yields
\begin{equation}
\begin{split}
D'' &+ D'\!\left[\frac{\mathcal{H}'}{\mathcal{H}}
    + 1 - 3w_{\rm DM}\right] \\
    &- D\!\left[
      \frac{3}{2}(1+w_{\rm DM})(1+3c_s^2)
      \Omega_{\rm DM}(a)\right. \\
    &\left.\qquad
      - \frac{c_s^2 k^2}{\mathcal{H}^2}
      - 3\frac{\mathcal{H}'}{\mathcal{H}}
        (c_s^2 - w_{\rm DM})
    \right] = 0,
\end{split}
\label{eq:growth_wDM}
\end{equation}
where here primes denote $d/d\ln a$ and
$\mathcal{H}' \equiv d\mathcal{H}/d\ln a$,
consistent with the notation of
equation~(\ref{eq:growth}). For $w_{\rm DM} =
c_s^2 = 0$, the coefficient of $D'$ reduces to
$\mathcal{H}'/\mathcal{H} + 1$ and the source
term reduces to $(3/2)\Omega_{\rm DM}(a)$,
recovering the $\Lambda$CDM growth
equation~(\ref{eq:growth}) with $\xi = 0$.

Equation~(\ref{eq:growth_wDM}) acquires a
$k$-dependence through the pressure term
$c_s^2 k^2/\mathcal{H}^2$, introducing a
characteristic Jeans scale
\begin{equation}
k_J(a) = \mathcal{H}(a)
    \sqrt{\frac{3(1+w_{\rm DM})(1+3c_s^2)
    \Omega_{\rm DM}(a)}{2c_s^2}},
\label{eq:kJeans}
\end{equation}
above which ($k > k_J$) pressure support suppresses
the growth of perturbations, while on scales
$k \ll k_J$ the pressure term is negligible and
the standard pressureless growth is recovered.
In the special case $c_s^2 = 0$, the $k$-dependence
vanishes entirely and equation~(\ref{eq:growth_wDM})
reduces to a scale-independent ODE, in which the
sole effect of $w_{\rm DM}$ is to modify the
background expansion $\mathcal{H}(a)$ and the
dilution of $\Omega_{\rm DM}(a)$.

\section{Methodology}
\label{methodology}

The cosmic structure formation in a given theory is most accurately modeled with N-body simulations, where the particles in the simulation are incrementally displaced from their initial positions according to the Poisson equation, the expansion rate of the universe and initial conditions. In this work, we will use the \texttt{pmwd} code~\cite{Li:2022bsu}, a differentiable cosmological particle-mesh N-body library.  In its native form,
\texttt{pmwd} supports $\Lambda$CDM and $w_0 w_a$CDM
cosmologies. To accommodate the IDE and $w_{\rm DM} \neq 0$ frameworks described
in Section~\ref{model}, we extended the code.

During each simulation, N-body particles discretize the
uniform distribution of matter at the beginning of cosmic
history at their Lagrangian positions $q$, from which they
then evolve by displacements $s$ to their later positions
$\mathbf{x} = q + s(q)$. The initial conditions are set
perturbatively when the linear density fluctuation is much
smaller than unity, using second-order Lagrangian
perturbation theory (2LPT)~\cite{Bouchet:1994xp}:
\begin{eqnarray}
    \mathbf{s} &=& D_1\mathbf{s}^{(1)} + D_2\mathbf{s}^{(2)},
    \nonumber \\
    \mathbf{p} &=& a^3 H\left[D_1'\mathbf{s}^{(1)} 
                 + D_2'\mathbf{s}^{(2)}\right],
\end{eqnarray}
where $\mathbf{p}$ is the canonical momentum conjugate to
$\mathbf{x}$, $H$ is the modified background expansion for the DM-DE coupled, and $D_{1,2}$ are the first-
and second-order growth factors whose evolution is governed
by equation~(\ref{eq:growth}).

Following~\cite{Li:2022bsu}, the growth equations
are written in terms of the suppression factors
\begin{equation}
    \mathrm{G}_m \equiv D_m/a^m,
\end{equation}
whose derivatives relate to those of $D_m$ as
\begin{align}
D_m        &= a^m \mathrm{G}_m, \\
D_m'       &= a^m\left(m\mathrm{G}_m
             + \mathrm{G}_m'\right), \\
D_m''      &= a^m\left(m^2\mathrm{G}_m
             + 2m\mathrm{G}_m'
             + \mathrm{G}_m''\right),
\end{align}
where primes denote $d/d\ln a$. For the first-order
growth factor one has $D_1 = a\mathrm{G}_1$,
$D_1' = a(\mathrm{G}_1 + \mathrm{G}_1')$, and
$D_1'' = a(\mathrm{G}_1 + 2\mathrm{G}_1' +
\mathrm{G}_1'')$.

The native growth integrator solves the standard equations governing structure growth in $\Lambda$CDM. We extended it to solve the coupled DM--DE system and the growth equations for $w_{\rm DM} \neq 0$. Both extensions recover the standard $\Lambda$CDM results in the appropriate limits, $\xi = 0$ and $w_{\rm DM} = 0$, respectively.

For the coupled DM--DE case, we define
\begin{equation}
\Omega_{\rm fac}^{\rm IDE} \equiv
    \frac{3}{2}\Omega_c\!\left(1 + \xi\,r_{xc}\right)
    - \frac{\mathcal{H}'}{\mathcal{H}}\,\xi\,r_{xc},
\label{eq:Omega_fac_IDE}
\end{equation}
and substituting $D_1 = a\mathrm{G}_1$ into
equation~(\ref{eq:growth}) yields
\begin{equation}
\begin{split}
\mathrm{G}_1''
    &+ \mathrm{G}_1'\!\left[
        \frac{\mathcal{H}'}{\mathcal{H}}
        + 3 + \xi\,r_{xc}\right] \\
    &+ \mathrm{G}_1\!\left[
        \frac{\mathcal{H}'}{\mathcal{H}}
        + 2 + \xi\,r_{xc}
        - \Omega_{\rm fac}^{\rm IDE}
        \right] = 0,
\end{split}
\label{G1}
\end{equation}
For the second-order factor,
$D_2 = a^2\mathrm{G}_2$, with
$D_2' = a^2(2\mathrm{G}_2 + \mathrm{G}_2')$ and
$D_2'' = a^2(4\mathrm{G}_2 + 4\mathrm{G}_2' +
\mathrm{G}_2'')$, the corresponding ODE reads
\begin{equation}
\begin{split}
\mathrm{G}_2''
    &+ \mathrm{G}_2'\!\left[
        \frac{\mathcal{H}'}{\mathcal{H}}
        + 5 + \xi\,r_{xc}\right] \\
    &+ \mathrm{G}_2\!\left[
        2\frac{\mathcal{H}'}{\mathcal{H}}
        + 6 + 2\xi\,r_{xc}
        - \Omega_{\rm fac}^{\rm IDE}
        \right] \\
    &= \frac{3}{2}\Omega_m\,\mathrm{G}_1^2.
\end{split}
\label{G2}
\end{equation}
The growth tables for
$\mathrm{G}_1$, $\mathrm{G}_2$ and their
derivatives are precomputed at the start of
each run and used throughout the 2LPT initial
conditions and the PM time-stepping.

Following the same procedure as in the coupled DM-DE case, we
substitute $D_1 = a\mathrm{G}_1$ into
equation~(\ref{eq:growth_wDM}) to obtain the
first-order suppression factor ODE. Defining
\begin{equation}
\Omega_{\rm fac} \equiv
    \frac{3}{2}(1+w_{\rm DM})(1+3c_s^2)\Omega_{\rm DM}
    - \frac{c_s^2 k^2}{\mathcal{H}^2}
    - 3\frac{\mathcal{H}'}{\mathcal{H}}(c_s^2-w_{\rm DM}),
\label{eq:Omega_fac_wDM}
\end{equation}
the first-order equation reads
\begin{equation}
\begin{split}
\mathrm{G}_1''
    &+ \mathrm{G}_1'\!\left[
        \frac{\mathcal{H}'}{\mathcal{H}}
        + 3 - 3w_{\rm DM}\right] \\
    &+ \mathrm{G}_1\!\left[
        \frac{\mathcal{H}'}{\mathcal{H}}
        + 2 - 3w_{\rm DM}
        - \Omega_{\rm fac}
        \right] = 0.
\end{split}
\label{eq:G1_wDM}
\end{equation}
The second-order
equation, obtained by substituting $D_2 = a^2\mathrm{G}_2$,
reads
\begin{equation}
\begin{split}
\mathrm{G}_2''
    &+ \mathrm{G}_2'\!\left[
        \frac{\mathcal{H}'}{\mathcal{H}}
        + 5 - 3w_{\rm DM}\right] \\
    &+ \mathrm{G}_2\!\left[
        2\frac{\mathcal{H}'}{\mathcal{H}}
        + 6 - 6w_{\rm DM}
        - \Omega_{\rm fac}
        \right] \\
    &= \frac{3}{2}(1+w_{\rm DM})
       \Omega_{\rm DM}\,\mathrm{G}_1^2.
\end{split}
\label{eq:G2_wDM}
\end{equation}

Note that both equations acquire
a scale dependence through the term
$c_s^2 k^2/\mathcal{H}^2$ contained in
$\Omega_{\rm fac}$, which is absent in the
$\Lambda$CDM and IDE cases. In the pressureless
limit $c_s^2 = 0$, this $k$-dependence vanishes
and the sole effect of $w_{\rm DM}$ enters through
the modified background $\mathcal{H}(a)$ and the
factor $(1+w_{\rm DM})$ in the source term of
equation~(\ref{eq:G2_wDM}).

To reduce the impact of cosmic variance on our power 
spectrum measurements, we employ the Fixed-and-Paired 
technique introduced by~\cite{Angulo:2016hjd}. In this 
approach, the amplitudes of the initial Fourier modes are 
fixed to $|\delta(\mathbf{k})| = \sqrt{P_\mathrm{lin}(k)}$, 
suppressing the stochastic scatter between realisations, 
while two simulations are run for each seed with phases 
$\phi$ and $\phi + \pi$. The average of each fixed-paired 
pair cancels the leading-order mode-coupling contribution 
to the variance, reducing the effective cosmic variance 
by more than an order of magnitude compared to standard 
random realisations~\cite{Angulo:2016hjd}.

Beyond the matter power spectrum $P_{\delta\delta}(k)$, 
we also measure the velocity divergence power spectra 
$P_{\theta\theta}(k)$ and $P_{\delta\theta}(k)$ directly 
from the simulations. The velocity divergence field is 
estimated in Fourier space as $\theta(\mathbf{k}) = 
i\mathbf{k}\cdot\mathbf{v}(\mathbf{k})$, where 
$\mathbf{v}$ is the particle velocity field assigned to 
a regular mesh via a volume-weighted Monte-Carlo Voronoi 
estimator~\cite{Bernardeau:1995en, Esposito:2024qlo}. This 
method assigns to each cell the velocity of its nearest 
N-body particle, evaluated at a set of quasi-random 
glass points drawn from a Halton low-discrepancy 
sequence~\cite{Halton:1960}, which provides a more 
uniform sampling of the volume than a Poisson 
distribution. The power spectra are then computed as
\begin{align}
P_{\theta\theta}(k) &= 
    \left\langle|\theta(\mathbf{k})|^2\right\rangle V,
\label{eq:Ptt_sim}
\\
P_{\delta\theta}(k) &= 
    \mathrm{Re}\left\langle
    \delta^*(\mathbf{k})\,\theta(\mathbf{k})
    \right\rangle V,
\label{eq:Pdt_sim}
\end{align}
where angle brackets denote spherical averages over 
shells of width $\Delta k$. In linear theory these 
reduce to equations~(\ref{eq:Ptt})--(\ref{eq:Pdt}), 
a relation we verify numerically as a consistency check 
of the pipeline. On non-linear scales, shell-crossing 
and vorticity generation break the linear 
relation~\cite{Pueblas:2008uv,Angulo:2021kes,Lepori:2026ptd,Sorensen:2025fuu,Esposito:2024qlo}, making $P_{\theta\theta}$ 
and $P_{\delta\theta}$ independent probes.

\subsubsection*{Simulation setup}

All simulations are run in two complementary
configurations, which we label Run-1 and Run-2,
each employing $512^3$ dark matter particles and a
PM mesh of $N_{\rm pm} = 1024^3$ cells. All runs start from
initial conditions generated at $z_{\rm ini} = 49$
using 2LPT with the growth factors appropriate to
each cosmological model, and are evolved to $z = 0$
in 60 time steps. Snapshots of particle positions
and velocities are saved at $z = 0$ and used for
all power spectrum and phase-space analyses
presented in this work.

Run-1 uses a periodic box of side $L_\mathrm{box} = 
512\,h^{-1}$Mpc, giving a particle mass of 
$m_\mathrm{part} \simeq 5.5\times10^{10}\,h^{-1}M_\odot$ 
and a reliable force scale of $k_\mathrm{rel} \simeq 
1.88\,h\,\mathrm{Mpc}^{-1}$. The large volume reduces 
cosmic variance at large scales ($k \lesssim 
0.1\,h\,\mathrm{Mpc}^{-1}$) and provides better 
sampling of the rare, massive end of the halo mass 
function.

Run-2 uses a periodic box of side $L_\mathrm{box} = 
256\,h^{-1}$Mpc, giving a particle mass of 
$m_\mathrm{part} \simeq 6.9\times10^{8}\,h^{-1}M_\odot$ 
and a reliable force scale of $k_\mathrm{rel} \simeq 
3.77\,h\,\mathrm{Mpc}^{-1}$. The smaller box provides 
80 times better mass resolution than Run-1, resolving 
haloes down to $M \sim 10^{10}\,h^{-1}M_\odot$, and extends the reliable $P(k)$ 
measurement to twice the wavenumber.

The key properties of both runs are summarised in
Table~\ref{tab:sims}. The two configurations are
complementary: Run-1 probes the large-scale regime
where the signals of coupled DM--DE and
$w_{\rm DM} \neq 0$ in $P_{\delta\delta}(k)$ first
depart from $\Lambda$CDM, while Run-2 resolves the
non-linear scales where the effects of the
DM--DE coupling parameter $\xi$ and of
$w_{\rm DM} \neq 0$ on the density and velocity
divergence spectra $P_{\delta\delta}$,
$P_{\theta\theta}$ and $P_{\delta\theta}$ are most
pronounced.

\begin{table}
\centering
\caption{Simulation parameters for Run-1 and Run-2.}
\label{tab:sims}
\begin{tabular}{lcc}
\hline\hline
Parameter & Run-1 & Run-2 \\
\hline
$N_\mathrm{part}$ & $512^3$ & $512^3$ \\
$L_\mathrm{box}$ [$h^{-1}$Mpc] & 512 & 256 \\
$m_\mathrm{part}$ [$h^{-1}M_\odot$] 
    & $5.5\times10^{10}$ & $6.9\times10^{8}$ \\
$N_\mathrm{pm}$ & $1024^3$ & $1024^3$ \\
$k_\mathrm{rel}$ [$h\,\mathrm{Mpc}^{-1}$] 
    & 1.88 & 3.77 \\
$z_\mathrm{ini}$ & 49 & 49 \\
\hline\hline
\end{tabular}
\end{table}

\begin{figure}
    \centering
    \includegraphics[width=\columnwidth]{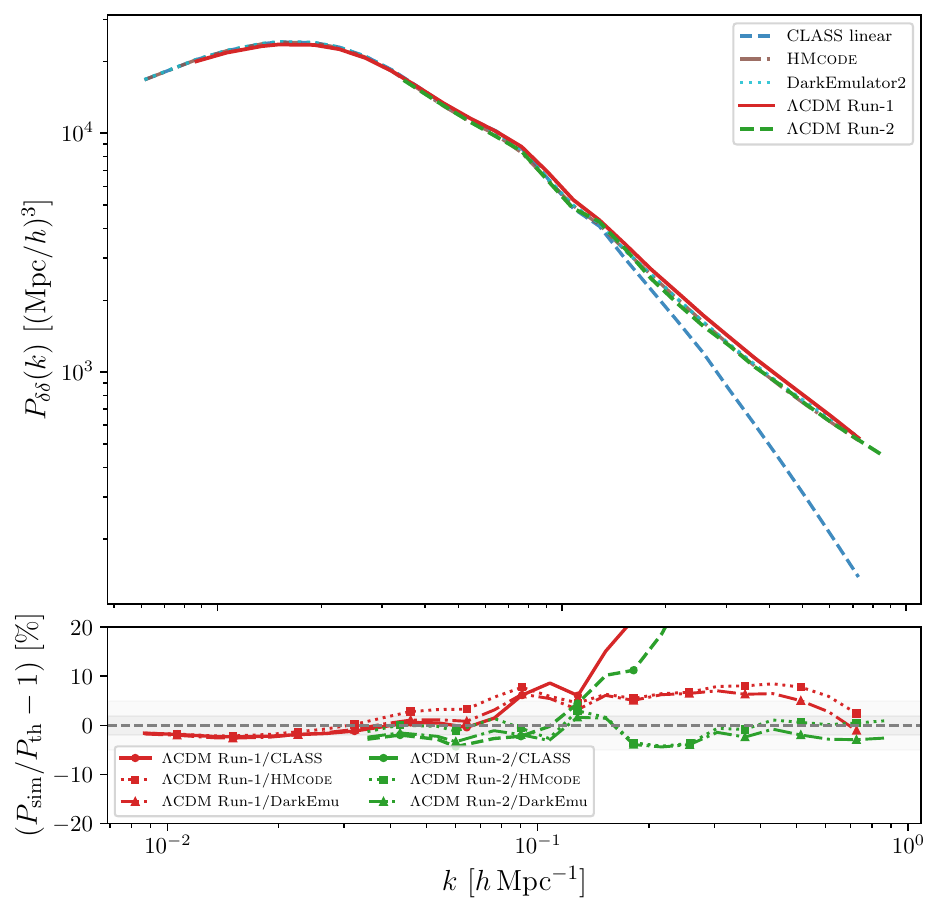}
    \caption{Non-linear matter power spectrum $P_{\delta\delta}(k)$ at $z=0$
for the $\Lambda$CDM model, comparing Run-1 and Run-2.
The dashed blue line shows the \texttt{CLASS} linear prediction, the dash-dotted
line the \texttt{HMcode-2020} non-linear model, and the dotted line the
\texttt{DarkEmulator2} emulator. The lower panel shows the relative difference
$(P_{\rm sim}/P_{\rm th} - 1)$ with respect to each theoretical reference.
The grey bands indicate the $2\%$ and $5\%$ precision levels.}
    \label{fig:Pdd}
\end{figure}

\begin{figure*}
    \centering
    \includegraphics[width=1.9\columnwidth]{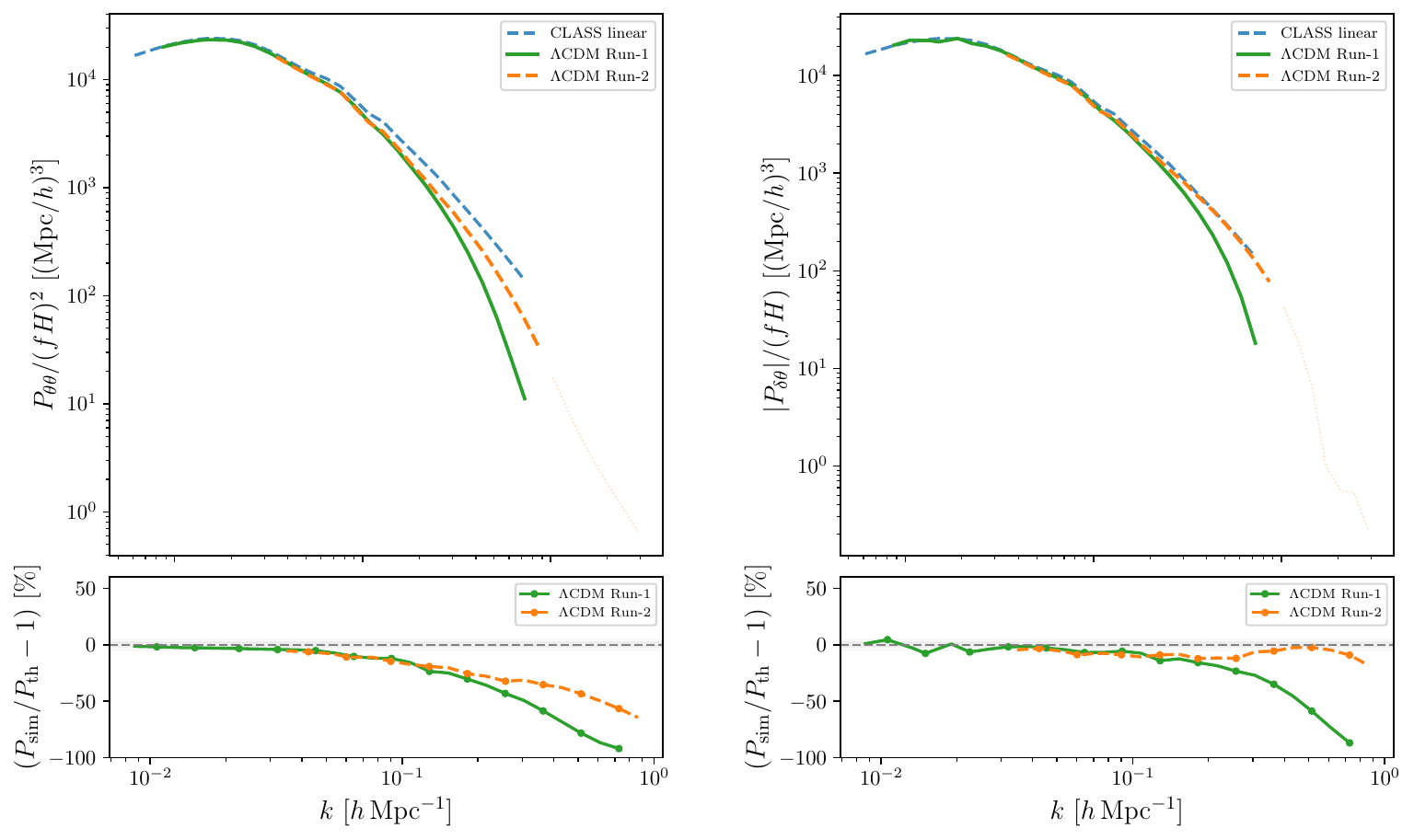}
    \caption{Normalised velocity divergence power spectra at $z=0$ for the $\Lambda$CDM model. Results are shown for Run-1 and Run-2. The dashed blue line shows the \texttt{CLASS} linear prediction. The lower panels show the relative differences with
respect to the linear theory. The grey bands indicate the $2\%$ and
$5\%$ precision levels.}
    \label{fig:Pvel}
\end{figure*}

\begin{figure*}
    \centering
    \includegraphics[width=1.9\columnwidth]{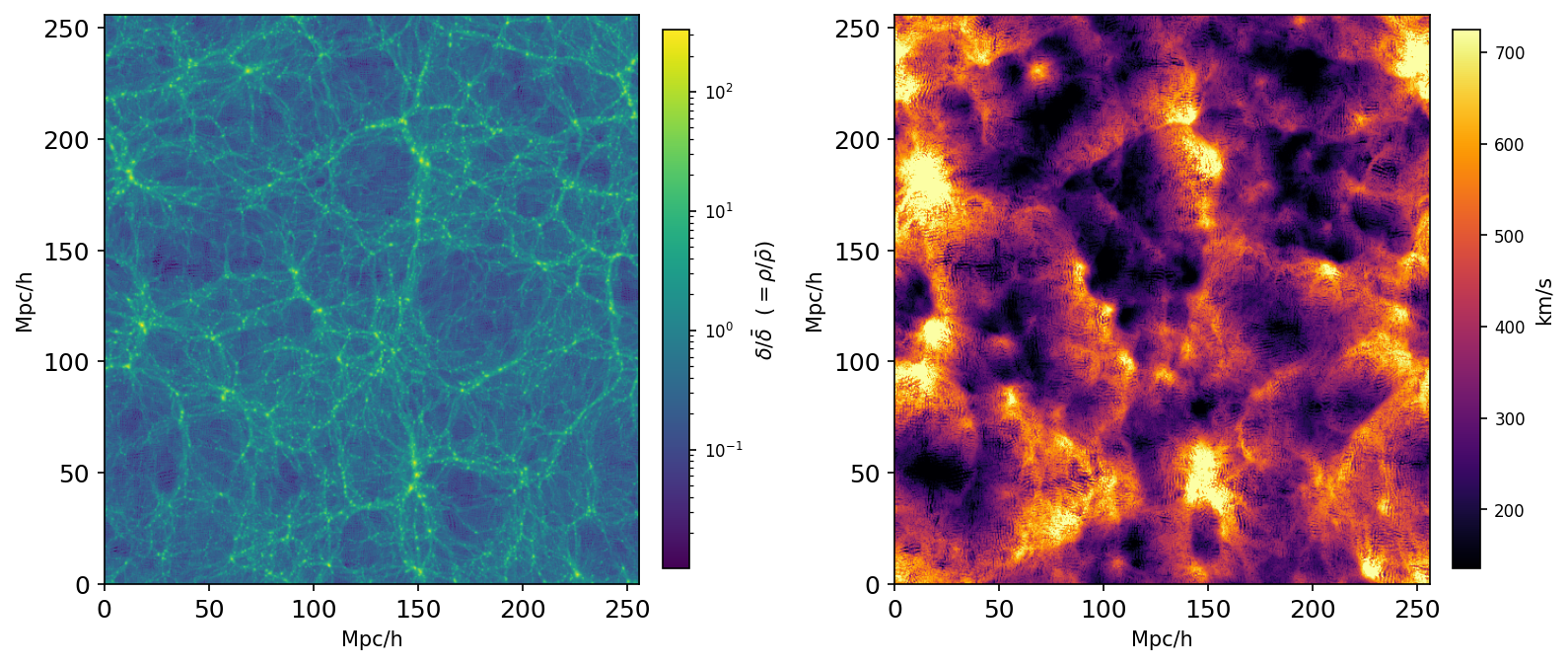}
    \caption{Projected dark matter density and velocity fields at $z=0$
for the $\Lambda$CDM model (Run-2). \textit{Left:} the density contrast field
$\delta/\bar{\delta} \equiv \rho/\bar{\rho}$, displayed on a logarithmic
colour scale, revealing the cosmic web of filaments, walls and voids.
\textit{Right:} the magnitude of the peculiar velocity field
$|\mathbf{v}|$ in km\,s$^{-1}$, estimated via a Monte-Carlo Voronoi
tessellation on a $512^3$ mesh with 16 glass points per cell.
High-velocity regions ($|\mathbf{v}| \gtrsim 500$\,km\,s$^{-1}$)
coincide with the nodes and filaments of the cosmic web visible in the
density field, while voids are characterised by coherent bulk flows
and low velocity amplitudes.}
    \label{fig:field_lcdm}
\end{figure*}

\section{Main results}
\label{results}

In what follows, we discuss our main results for the matter and velocity-divergence power spectra, $P_{\delta\delta}$, $P_{\theta\theta}$, and $P_{\delta\theta}$, extracted from the simulation snapshots up to $k = 1.0\,h\,{\rm Mpc}^{-1}$. Results at smaller scales are discussed only for the simulations with $w_{\rm DM} \neq 0$.

This range extends sufficiently into the non-linear regime to serve the primary goal of this work: identifying the scales that provide the greatest discriminatory power between the standard $\Lambda$CDM model and the non-standard scenarios considered here. The results presented in this section are intended to guide future theoretical modelling efforts and the development of accurate analytical prescriptions that can be confronted with data from ongoing and forthcoming spectroscopic surveys, such as Euclid and DESI. In particular, these results may provide valuable insights for future full-shape analyses of galaxy clustering over the range $k < 1.0\,h\,{\rm Mpc}^{-1}$.

\subsubsection*{Validation in $\Lambda$CDM}

Before analysing the IDE and $w_{\rm DM} \neq 0$ 
scenarios, we validate our modified pipeline against 
the standard $\Lambda$CDM model, adopting the 
Planck-CMB best-fit cosmological 
parameters~\cite{Planck:2018vyg}. 

Figures~\ref{fig:Pdd} and~\ref{fig:Pvel} show the matter 
and velocity divergence power spectra at $z=0$ for the $\Lambda$CDM 
model, comparing Run-1 and Run-2 against theoretical references.

Figure~\ref{fig:Pdd} shows that both runs are in excellent 
agreement with the \texttt{CLASS} linear~\cite{Diego_Blas_2011} prediction on large scales, with deviations below $2\%$, 
confirming that the 2LPT initial conditions and the background 
expansion are correctly implemented. On intermediate scales 
($0.05 \lesssim k \lesssim 0.2\,h\,{\rm Mpc}^{-1}$), both runs 
depart from the linear prediction as expected, with the non-linear 
enhancement of power well captured by both \textsc{HMcode-2020} ~\cite{Mead:2020vgs} and 
\texttt{DarkEmulator2}~\cite{Nishimichi:2018etk,Miyatake:2020uhg}, to which the simulations agree below $5\%$. 

At smaller scales ($k \gtrsim 0.2\,h\,{\rm Mpc}^{-1}$), the two 
runs begin to diverge from each other. Run-1 departs from the 
theoretical references at $k \gtrsim 0.4\,h\,{\rm Mpc}^{-1}$, 
approaching its reliable force scale of 
$k_{\rm rel} = 1.88\,h\,{\rm Mpc}^{-1}$. Run-2, with its smaller 
box and correspondingly finer particle spacing, maintains agreement 
with \textsc{HMcode-2020} and \texttt{DarkEmulator2} to smaller scales, 
reaching $k \sim 1\,h\,{\rm Mpc}^{-1}$ with deviations below $4\%$. 
The upturn visible in Run-1 at $k \gtrsim 0.8\,h\,{\rm Mpc}^{-1}$ 
is a known numerical artefact of the PM force softening near 
$k_{\rm rel}$, and not a physical signal. These results confirm 
that the pipeline is correctly validated against established 
$\Lambda$CDM predictions.

Figure~\ref{fig:Pvel} shows the normalised velocity divergence 
spectra $P_{\theta\theta}/(fH)^2$ and $|P_{\delta\theta}|/(fH)$, 
which in linear theory reduce to $P_{\delta\delta}^{\rm lin}(k)$.
Both spectra follow the \texttt{CLASS} linear prediction closely on large 
scales, with deviations 
below $5\%$, confirming that the velocity field estimator is 
correctly normalised and that the growth rate $f$ is accurately 
reproduced by the simulations.

On non-linear scales, the velocity spectra are more sensitive to 
resolution effects than $P_{\delta\delta}$, because the velocity 
field is estimated via a Monte-Carlo Voronoi tessellation that 
becomes noisier in under-resolved regions. This is clearly visible 
in the lower panels of Figure~\ref{fig:Pvel}: Run-1 begins to 
deviate significantly from the linear prediction at 
$k \gtrsim 0.1\,h\,{\rm Mpc}^{-1}$, reaching deviations of order 
$-50\%$ at $k \sim 0.5\,h\,{\rm Mpc}^{-1}$ and $-100\%$ near 
$k \sim 1\,h\,{\rm Mpc}^{-1}$. Run-2, with its finer mass resolution 
and smaller box, follows the linear prediction to smaller scales 
and shows significantly reduced deviations across the range 
$0.1 \lesssim k \lesssim 0.5\,h\,{\rm Mpc}^{-1}$, before also 
departing at $k \gtrsim 0.5\,h\,{\rm Mpc}^{-1}$. The good agreement we find on fully linear scales, where a direct comparison with a well-established theoretical prediction is possible, demonstrates the consistency and reliability of our simulations.

This behaviour reflects a fundamental difference between the density 
and velocity fields: whereas $P_{\delta\delta}$ is dominated by the 
high-density nodes of the cosmic web where the PM force is well 
resolved even in Run-1, the velocity field receives contributions 
from the low-density voids and filaments where the particle sampling 
is sparse and the Voronoi tessellation is less accurate. Increasing 
$L_{\rm box}$ at fixed $N_{\rm part}$, as in Run-1 relative to 
Run-2, increases the particle mass and reduces the number density 
of tracers, directly degrading the accuracy of the velocity field 
reconstruction at intermediate and small scales. Run-2 therefore 
provides a substantially more reliable measurement of 
$P_{\theta\theta}$ and $P_{\delta\theta}$ in the non-linear regime, 
and we adopt it as our primary reference for the velocity field 
analysis in the non-standard scenarios discussed below.
These results are consistent with previous findings in the 
literature, which established that 
accurate velocity field measurements require higher mass resolution. Figure \ref{fig:field_lcdm} shows the projected DM density and velocity fields from the Run-2.

\begin{figure*}
    \centering
    \includegraphics[width=1.6\columnwidth]{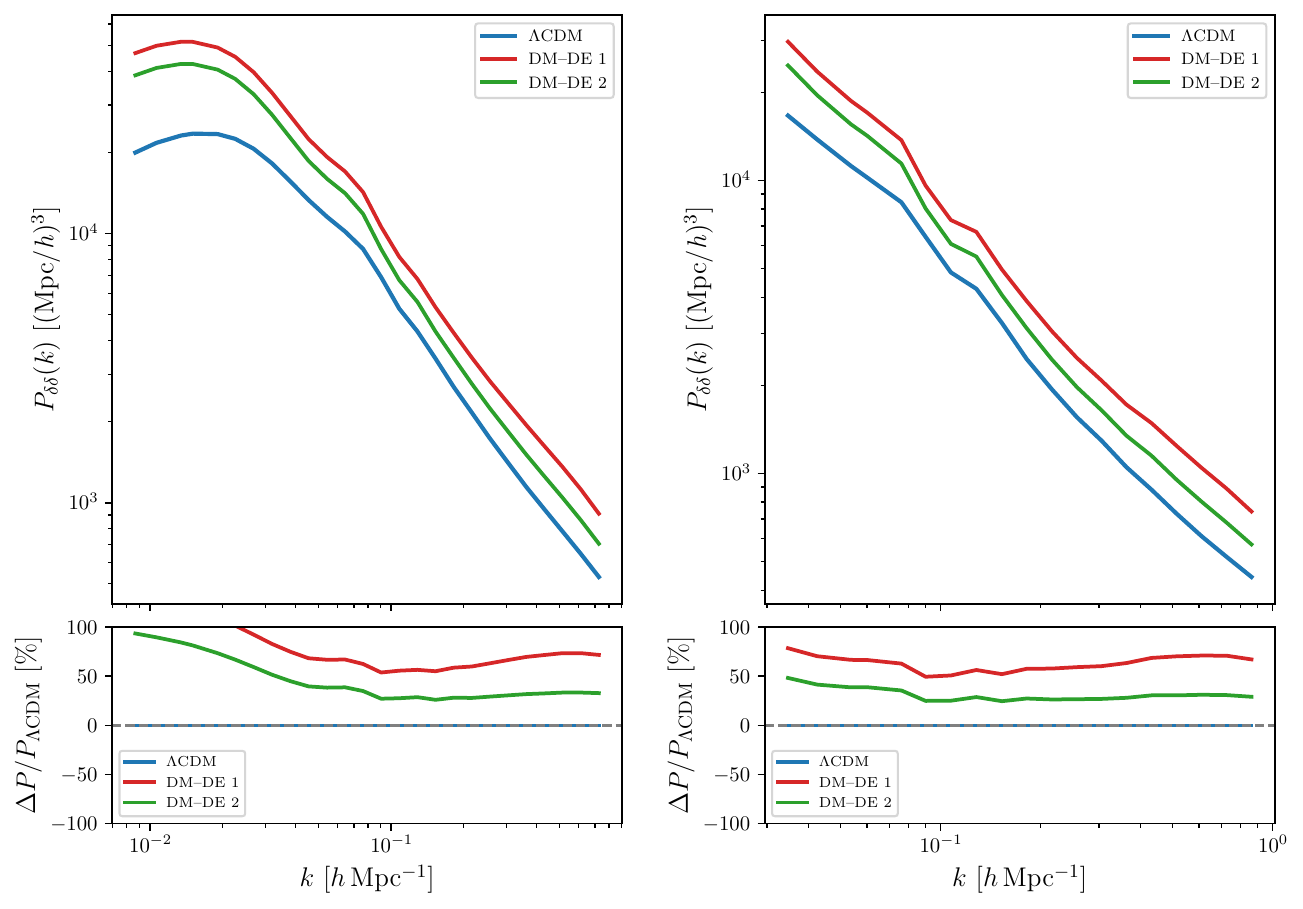}
    \caption{Non-linear matter power spectrum $P_{\delta\delta}(k)$ at $z=0$
for the $\Lambda$CDM model and the two coupled DM--DE scenarios,
DM--DE~1 and DM--DE~2. The left and right
panels correspond to Run-1  and
Run-2, respectively. The lower panels
show the relative difference
$\Delta P/P_{\Lambda{\rm CDM}} \equiv
(P_{\rm model} - P_{\Lambda{\rm CDM}})/P_{\Lambda{\rm CDM}}$
with respect to the $\Lambda$CDM spectrum.}
    \label{fig:IDE_dd}
\end{figure*}

\begin{figure*}
    \centering
    \includegraphics[width=1.6\columnwidth]{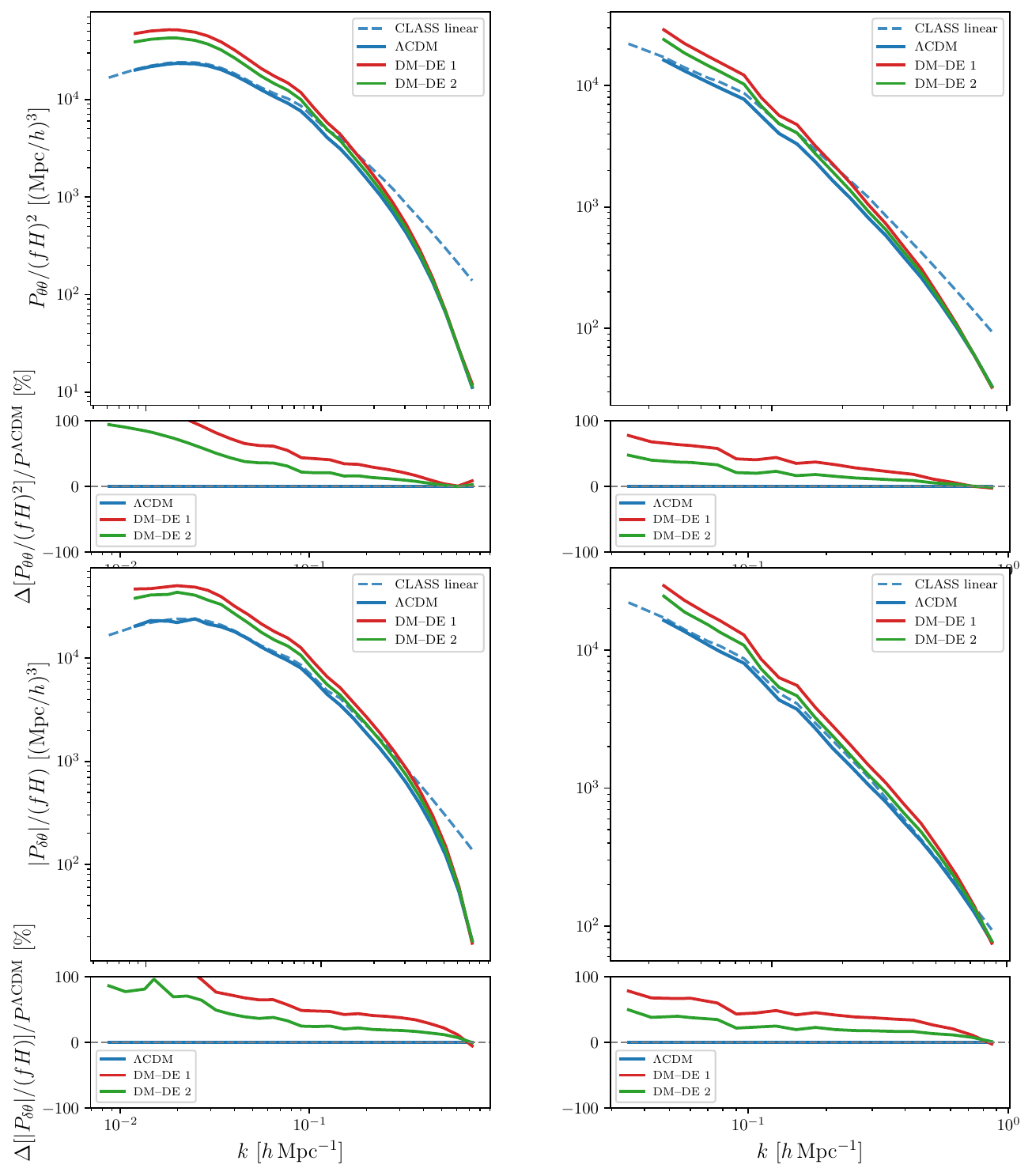}
    \caption{Normalised velocity divergence power spectra at $z=0$
for the $\Lambda$CDM model and the two coupled DM--DE scenarios,
DM--DE~1 ($\xi=-0.1$) and DM--DE~2 ($\xi=-0.05$). Results are shown for Run-1 and Run-2.
The upper two panels show $P_{\theta\theta}/(fH)^2$ and the lower
two panels show $|P_{\delta\theta}|/(fH)$, both in units of
$(\rm Mpc/\mathit{h})^3$.
The dashed blue line shows the CLASS linear prediction
$P_{\delta\delta}^{\rm lin}(k)$, to which both normalised ratios
reduce in the linear regime.
The lower sub-panels show the relative difference
$\Delta P / P_{\Lambda{\rm CDM}}$ with respect to the
$\Lambda$CDM spectrum.}
    \label{fig:IDE_vel}
\end{figure*}

\begin{figure*}
    \centering
    \includegraphics[width=1.9\columnwidth]{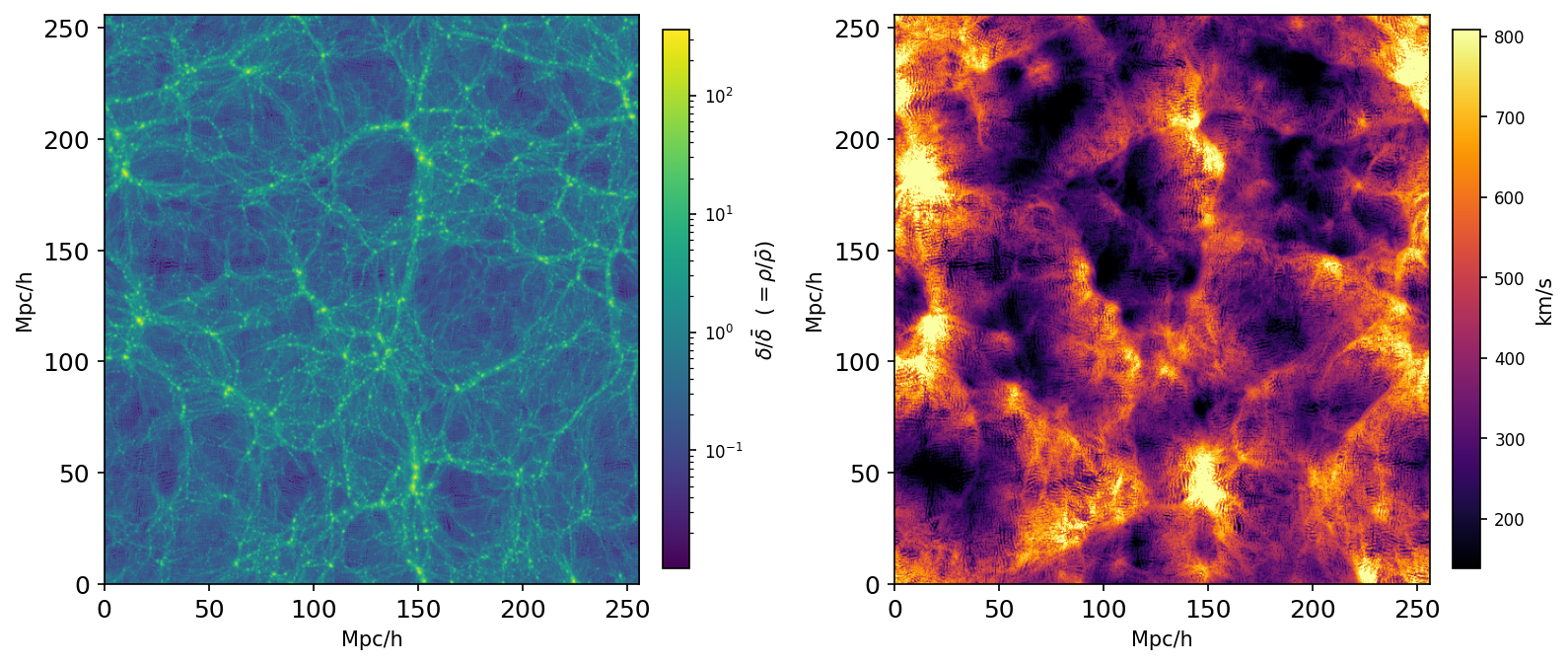}
\caption{Same as Fig.~\ref{fig:field_lcdm}, but for the coupled
DM--DE simulation DM--DE~1.}
    \label{fig:field_IDE}
\end{figure*}

\subsubsection*{Coupled DM--DE}

To quantify the impact of the DM--DE interaction
on the non-linear matter and velocity power spectra,
we adopt fiducial cosmological parameters grounded
in current observational constraints. As baseline
values for all standard parameters
($\Omega_m$, $\Omega_b$, $h$, $n_s$, $\sigma_8$),
we take the constraints reported in Table~II
of~\cite{Silva:2025hxw}, derived from the joint analysis of
CMB and BAO DESI DR2 data within the coupled
DM--DE framework. For the coupling parameter $\xi$,
we investigate two values that bracket the current
observational limits: $\xi = -0.1$ and
$\xi = -0.05$. These define our two coupled DM--DE
simulation runs, which we label \textbf{DM--DE~1} and
\textbf{DM--DE~2}, respectively.

\textit{It is important to emphasise that the baseline
cosmological parameters
($\Omega_m$, $\Omega_b$, $h$, $n_s$, $\sigma_8$)
take different values between the $\Lambda$CDM and
coupled DM--DE runs. This is a deliberate choice: rather
than isolating the effect of $\xi$ alone by varying
only the coupling while keeping all other parameters
fixed, we aim to compare the models as they are
actually constrained by current data. In this way,
our simulations reflect the full parameter
degeneracies present in observational analyses and
provide a more realistic assessment of how the
coupled DM--DE scenario differs from $\Lambda$CDM
across a range of scales. Our goal is to identify
which scales carry the most discriminating power
between these scenarios, thereby informing the
design of future observational tests capable of
decisively distinguishing between them.}

Figures~\ref{fig:IDE_dd} and~\ref{fig:IDE_vel} show the
non-linear matter power spectrum $P_{\delta\delta}(k)$ and the
normalised velocity divergence spectra $P_{\theta\theta}/(fH)^2$
and $|P_{\delta\theta}|/(fH)$ at $z=0$, for the $\Lambda$CDM
model and the two coupled DM--DE scenarios, DM--DE~1
and DM--DE~2, for both simulation runs.

Figure~\ref{fig:IDE_dd} reveals that both DM--DE scenarios
produce significant and scale-dependent deviations from
$\Lambda$CDM across the entire range of scales probed,
$0.01 \lesssim k \lesssim 1\,h\,{\rm Mpc}^{-1}$.
The relative differences $\Delta P/P_{\Lambda{\rm CDM}}$ reach
values of order $50$--$100\%$ at the largest scales accessible
to Run-1, and remain substantial throughout the non-linear
regime probed by Run-2.

On large scales ($k \lesssim 0.05\,h\,{\rm Mpc}^{-1}$), the
suppression of $P_{\delta\delta}$ relative to $\Lambda$CDM is
most pronounced. This is the regime dominated by the linear
growth factor $D(a)$, which is directly modified by the
DM--DE coupling through equation~(\ref{eq:growth}): a negative
$\xi$ transfers energy from DM to DE, reducing the effective
gravitational source and slowing the growth of density
perturbations. The effect is stronger for DM--DE~1
($\xi=-0.1$) than for DM--DE~2 ($\xi=-0.05$), as expected
from the larger magnitude of the coupling. It is important
to note that part of the signal at large scales also reflects
the different baseline cosmological parameters adopted for each model,
constrained by CMB and BAO DESI DR2 data within the
respective framework.

On non-linear scales ($0.1 \lesssim k \lesssim 1\,h\,{\rm Mpc}^{-1}$), the
behaviour of $\Delta P/P_{\Lambda{\rm CDM}}$ changes
character. The relative difference becomes less negative
or even changes sign at certain scales, reflecting the
interplay between the suppressed linear growth and the
modified non-linear mode coupling. The DM--DE coupling
alters the halo mass function and the concentration-mass
relation, which in turn modifies the amplitude and shape
of $P_{\delta\delta}$ at these scales \cite{Zhai:2026uwr}. The consistency of the relative differences between Run-1 and Run-2 within their common $k$-range ($0.05 \lesssim k \lesssim 0.9\,h\,{\rm Mpc}^{-1}$) provides an important cross-validation of the pipeline and confirms that the IDE signatures are robust against resolution effects. Similar to the findings reported in \cite{Zhai:2026uwr}, we observe that, for scales below $1\,h\,{\rm Mpc}^{-1}$, the matter power spectrum amplitude in the coupled DM-DE scenario falls below the $\Lambda$CDM prediction in Run-2. This behavior points to a non-trivial modification of the structure formation process and motivates further theoretical investigations into the physical mechanisms governing the matter power spectrum on small scales. 

Figure~\ref{fig:IDE_vel} shows the normalised velocity
divergence spectra, which carry complementary and in some
respects more discriminating information than
$P_{\delta\delta}$ alone. In linear theory both
$P_{\theta\theta}/(fH)^2$ and $|P_{\delta\theta}|/(fH)$
reduce to $P_{\delta\delta}^{\rm lin}(k)$, providing a
baseline against which non-linear and model-dependent
effects can be cleanly identified.

The upper panels of Figure~\ref{fig:IDE_vel} show that,
on large scales, the DM--DE scenarios depart significantly
from $\Lambda$CDM in both $P_{\theta\theta}/(fH)^2$ and
$|P_{\delta\theta}|/(fH)$. The deviations are qualitatively
similar to those in $P_{\delta\delta}$, reflecting the common
origin in the modified growth factor $D(a)$ and growth rate
$f(a) = d\ln D/d\ln a$. However, the amplitude of the
relative differences in the velocity spectra differs from
that in $P_{\delta\delta}$, because $f(a)$ and $H(a)$ are
independently modified by the dark coupling. This means that even if $P_{\delta\delta}$
were held fixed, the velocity spectra would still deviate
from $\Lambda$CDM through the modified $fH$ factor, making
them sensitive probes of the DM--DE interaction that are
partially independent of the density field.

On non-linear scales the relative differences in the
velocity spectra become large and negative, reaching
$\sim -100\%$ near $k \sim 1\,h\,{\rm Mpc}^{-1}$ for
Run-1. This behaviour is partly physical, shell-crossing
and vorticity generation progressively decouple
$\theta$ from $\delta$ at small scales, breaking the
linear relation~\cite{Pueblas:2008uv}, and partly a
resolution effect, as the Monte-Carlo Voronoi velocity
estimator becomes noisier in under-sampled regions. As
established in the $\Lambda$CDM validation, Run-2
provides a substantially more reliable measurement of the
velocity spectra in the non-linear regime, and it is
Run-2 that should be used for quantitative comparison
with theoretical models at $k \gtrsim 0.1\,h\,{\rm Mpc}^{-1}$.

Comparing $P_{\theta\theta}$ and $P_{\delta\theta}$
directly, the coupled DM--DE signal is visible at
broadly similar amplitudes in both spectra, as expected
in the linear regime where the two are proportional.
The slight differences in their non-linear behaviour
arise from the distinct sensitivity of $\theta$ to
vorticity and from the cross-correlation structure
between density and velocity fields. Together, the
three statistics $P_{\delta\delta}$,
$P_{\theta\theta}$, and $P_{\delta\theta}$ provide
a more complete characterisation of the DM--DE
coupling signal than any single observable alone,
and their joint measurement from forthcoming
spectroscopic surveys such as Euclid and DESI will
substantially improve the discriminating power between
these scenarios. 

These effects suggest that $P_{\theta\theta}$ and $P_{\delta\theta}$ may contain valuable information for future observational tests that rely on these quantities as key inputs. To the best of our knowledge, this is the first investigation of such effects for this class of models. Figure \ref{fig:field_IDE} shows the projected dark matter density and velocity fields at $z = 0$ for the coupled DM-DE 1 model in Run-2.

\begin{figure}
    \centering
    \includegraphics[width=\columnwidth]{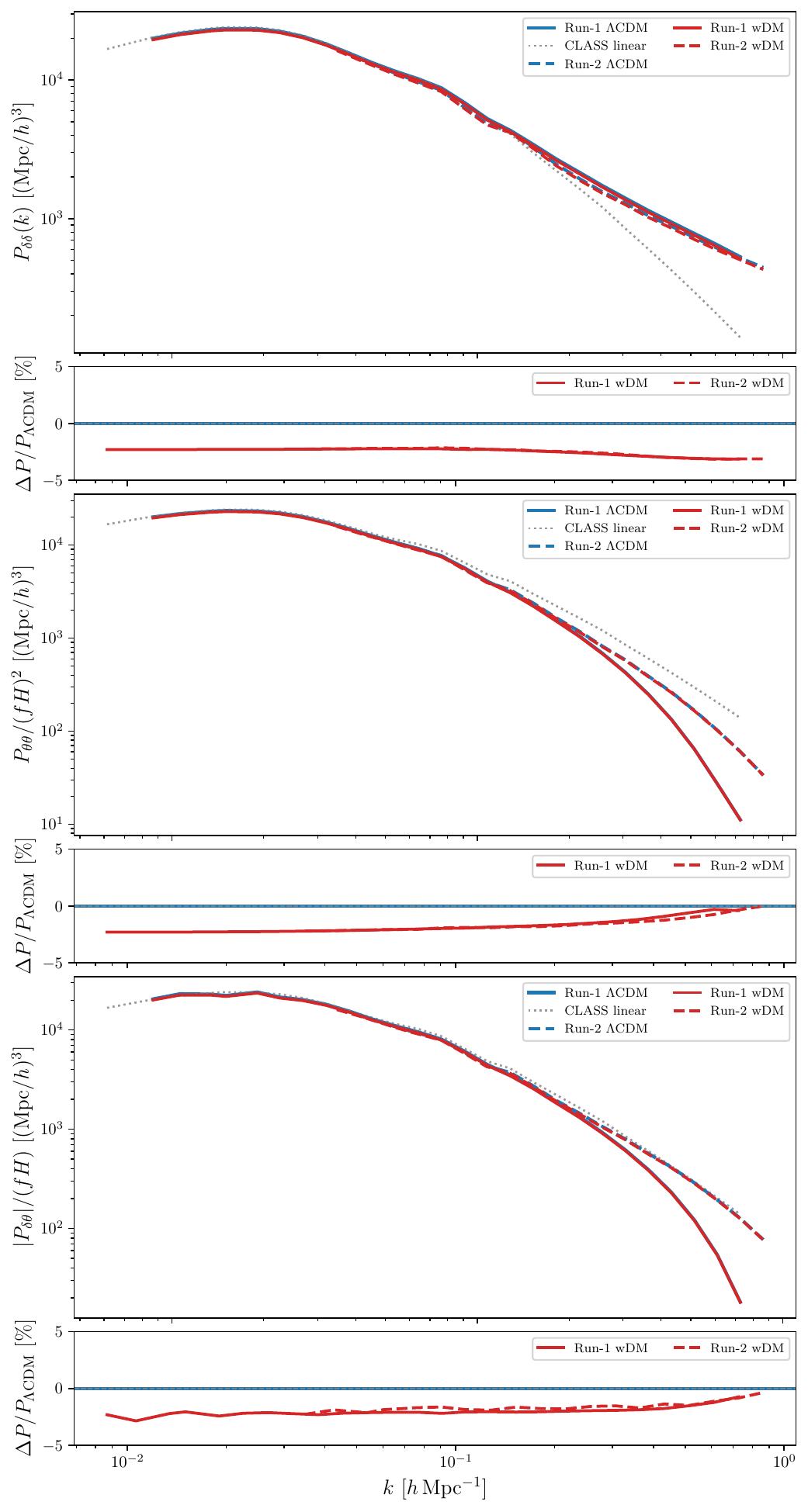}
    \caption{Matter and velocity divergence power spectra at $z=0$ for
the $\Lambda$CDM model and the $w_{\rm DM}\neq 0$ scenario
for Run-1 and Run-2. From top to bottom: matter power spectrum $P_{\delta\delta}(k)$,
normalised velocity divergence spectrum $P_{\theta\theta}/(fH)^2$,
and normalised cross-spectrum $|P_{\delta\theta}|/(fH)$, all in
units of $(\rm Mpc/\mathit{h})^3$. The dotted grey line shows the
CLASS linear prediction. The lower sub-panels show the relative
difference $\Delta P/P_{\Lambda{\rm CDM}}$ with respect to the
$\Lambda$CDM spectrum. The baseline cosmological parameters are
identical between the two scenarios; only $w_{\rm DM}$ differs.}
    \label{fig:Pdd_wDM}
\end{figure}

\begin{figure*}
    \centering
    \includegraphics[width=1.9\columnwidth]{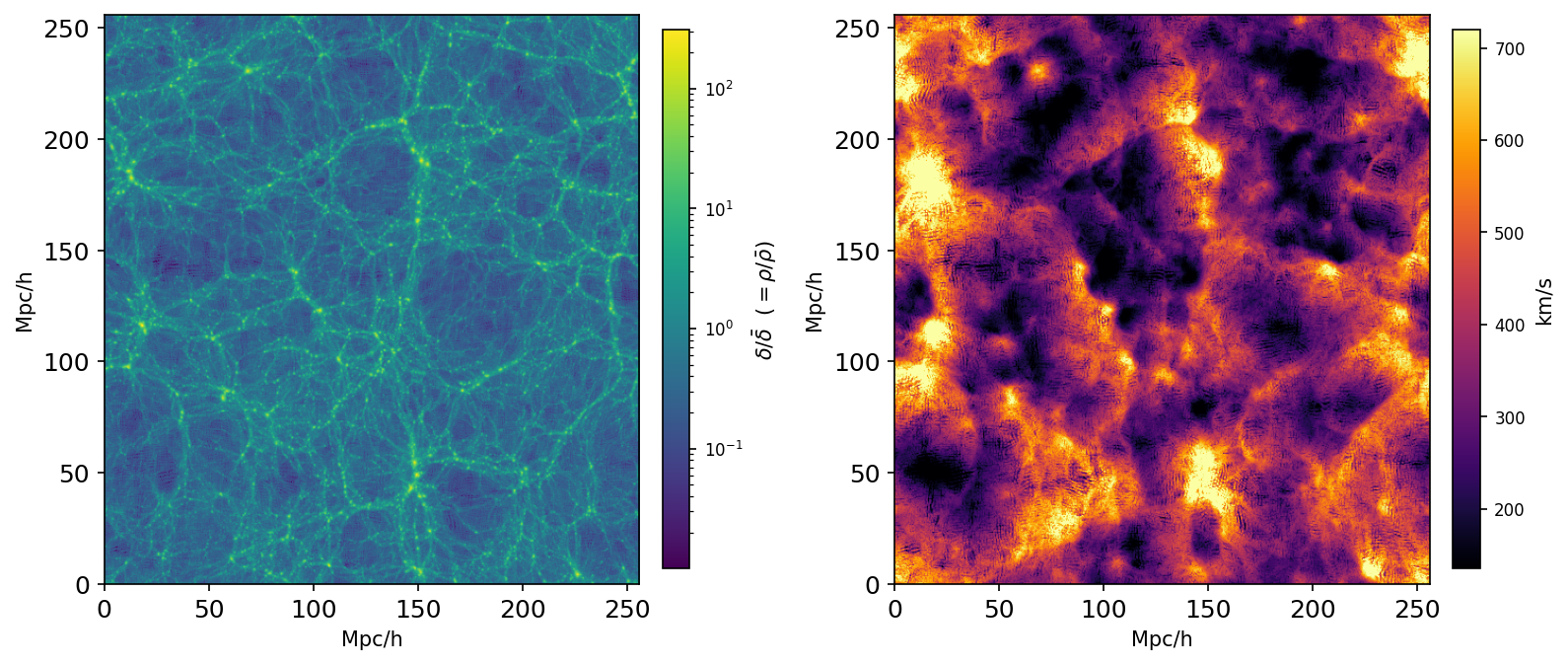}
\caption{Same as Fig.~\ref{fig:field_lcdm}, but for the $w_{\rm DM}\neq 0$ scenario.}
    \label{fig:field_wDM}
\end{figure*}

\subsubsection*{Simulations in the $w_{\rm DM} \neq 0$ Framework}

For dark matter with a non-zero equation-of-state parameter, we perform both Run-1 and Run-2, adopting $w_{\rm DM} = 10^{-3}$ and $c_s^2 = 10^{-6}$. These values are consistent with existing observational constraints from CMB measurements and geometrical probes (e.g., \cite{Kumar:2019gfl}), and can be regarded as representative upper bounds on the allowed departures from the standard pressureless DM scenario. The results can be straightforwardly extrapolated to smaller values of $w_{\rm DM}$, without loss of generality.

Figure~\ref{fig:Pdd_wDM} shows the matter and velocity divergence
power spectra at $z=0$ for the $w_{\rm DM}\neq 0$ scenario compared to $\Lambda$CDM.
Unlike the coupled DM--DE case, the baseline cosmological
parameters ($\Omega_m$, $h$, $\sigma_8$) are held identical
between the two models, so any difference in the spectra can be
attributed exclusively to the non-zero DM equation of state.

The relative differences $\Delta P_{\delta\delta}/P_{\Lambda{\rm CDM}}$
are remarkably small across all scales probed, remaining within
$\pm 5\%$ for both Run-1 and Run-2. This is physically expected:
in the pressureless limit, the sole effect of $w_{\rm DM}$
is to modify the background expansion through the altered dilution
of the CDM density (eq.~\ref{eq:rho_wDM}), and through the consequent change
in the growth factor $D(a)$ (equation~\ref{eq:growth_wDM}).
For $w_{\rm DM}= 10^{-3}$, the CDM density dilutes slightly faster
than in $\Lambda$CDM, reducing the gravitational source and
mildly suppressing the growth of structure. The effect is
cumulative over the expansion history but small in amplitude
for $|w_{\rm DM}| \ll 1$.

On large scales ($k \lesssim 0.05\,h\,{\rm Mpc}^{-1}$), the
suppression in $P_{\delta\delta}$ is most pronounced, of order
$2$--$4\%$, reflecting the scale-independent modification of
the linear growth factor. On non-linear scales
($k \gtrsim 0.1\,h\,{\rm Mpc}^{-1}$), the relative difference
becomes slightly less negative, as the enhanced non-linear
mode coupling partially compensates the reduced linear growth.
The consistency between Run-1 and Run-2 across their common
$k$-range confirms that the signal is physical and not a
numerical artefact.

The velocity divergence spectra $P_{\theta\theta}/(fH)^2$ and
$|P_{\delta\theta}|/(fH)$ show qualitatively similar behaviour
to $P_{\delta\delta}$, with relative differences confined within
$\pm 5\%$. This is consistent with the linear relation
$P_{\theta\theta}/(fH)^2 = P_{\delta\theta}/(fH) =
P_{\delta\delta}$ holding to a good approximation at the scales
where the $w_{\rm DM}$ signal is detectable. The agreement between
the two normalised velocity spectra and $P_{\delta\delta}$ at
large scales confirms that the growth rate $f(a) = d\ln D/d\ln a$
is also modified by $w_{\rm DM}$ consistently with the density
field, as expected from the common origin of both in the
growth equation~(\ref{eq:growth_wDM}).

At nonlinear scales ($k \gtrsim 0.3\,h\,{\rm Mpc}^{-1}$), the velocity spectra progressively depart from $P_{\delta\delta}$ as shell crossing and vorticity generation lead to a breakdown of the linear density–velocity relation, as expected. Run-2 provides more reliable measurements in this regime owing to its higher mass resolution. Figure \ref{fig:field_wDM} shows the projected dark matter density and velocity fields at $z=0$ for Run-2.

\section{Final Remarks}
\label{final}

In this work, we have investigated the impact of two
extensions of the standard $\Lambda$CDM
scenario on the statistical properties of the matter
and velocity divergence fields, from the linear regime
to non-linear scales up to $k = 1\,h\,{\rm Mpc}^{-1}$.
The two scenarios considered, coupled DM-DE interactions, parametrised by the
coupling $\xi$, and a non-zero dark matter equation of
state $w_{\rm DM}$, both modify the background
expansion history $H(a)$ and the growth of structure
$D(a)$, leaving distinctive imprints on the matter power
spectrum $P_{\delta\delta}(k)$ and on the velocity
divergence spectra $P_{\theta\theta}(k)$ and
$P_{\delta\theta}(k)$. 

For the coupled DM--DE scenario, adopting cosmological
parameters constrained by CMB and BAO DESI DR2 data within
the respective framework, we find significant and
scale-dependent deviations from $\Lambda$CDM. These
deviations are most pronounced at large scales, where the
linear growth factor $D(a)$ is directly suppressed by the
dark sector coupling, and remain substantial throughout the
non-linear regime. The velocity divergence spectra
$P_{\theta\theta}$ and $P_{\delta\theta}$ provide
complementary and partially independent sensitivity to
the coupling $\xi$.  For the $w_{\rm DM} \neq 0$ scenario, we find
systematic but remarkably small deviations from $\Lambda$CDM,
confined within $\pm 5\%$ across all scales probed, fully
consistent between the density and velocity fields. 

Our results reveal the non-trivial nature of nonlinear 
effects beyond the standard $\Lambda$CDM cosmology, and 
highlight the theoretical challenge of developing accurate 
analytical models capable of describing these simulations. 
Whether through the EFTofLSS or halo model approaches, such 
theoretical frameworks must be extended and calibrated 
against nonlinear simulations before these non-standard 
dark matter scenarios can be meaningfully confronted with 
current and forthcoming observational data.

Looking ahead, our results motivate some directions
for future work. First, extending the simulation suite
to a grid in the parameter space $(\xi, w_{\rm DM})$
would enable the construction of emulators for
$P_{\delta\delta}$, $P_{\theta\theta}$ and
$P_{\delta\theta}$ in non-standard cosmologies,
providing the theoretical machinery needed for
likelihood analyses of forthcoming spectroscopic data.
Second, the inclusion of massive neutrinos, which are
known to suppress structure growth on small scales and
may partially mimic the effects of $\xi$ or $w_{\rm DM}$,
will be necessary for a complete and unbiased modelling
of the non-linear regime. Together, these developments will
allow the models investigated here to be tested against
the full statistical power of ongoing and forthcoming
surveys such as DESI~\cite{DESI:2024jxi,Guandalin:2026yom} and Euclid~\cite{Euclid:2025diy}
on both linear and non-linear scales, with the
potential to decisively constrain or detect
departures from the standard cosmological model in
the dark sector.

\section*{Acknowledgements}
\noindent 
I acknowledge financial support from the Conselho Nacional de Desenvolvimento Científico e Tecnológico (CNPq, National Council for Scientific and Technological Development) through Grant No. 304306/2022-3, as well as partial financial support from the Fundação de Amparo à Pesquisa do Estado do Rio Grande do Sul (FAPERGS, Research Support Foundation of the State of Rio Grande do Sul) through Grant Nos. 23/2551-0000848-3 and 25/2551-0002612-1.

\section*{Data Availability}
The simulation snapshots, power spectra outputs, and
simulation code used in this work will be made
publicly available upon publication of this article.

\bibliography{main}

\end{document}